# Nanosculpted 3D helices of a magnetic Weyl semimetal with switchable nonreciprocity

Max T. Birch[1], Yukako Fujishiro[1], Ilya Belopolski[1], Masataka Mogi[2], Yi-Ling Chiew[1],

Xiuzhen Yu[1], Naoto Nagaosa[1,3], Minoru Kawamura[1], Yoshinori Tokura[1,2,4]

[1]*RIKEN Center for Emergent Matter Science (CEMS), Wako, Saitama 351-0198, Japan.*

[2]*Department of Applied Physics, The University of Tokyo, Bunkyo-ku, Tokyo 113-8656, Japan.*

[3]*RIKEN Fundamental Science Program, Wako, Saitama 351-0198, Japan.*

[4]*Tokyo College, The University of Tokyo, Bunkyo-ku, Tokyo 113-8656, Japan.*

**The emergent properties of materials are defined by the symmetries of their underlying atomic, spin and charge order. The explorations of symmetry breaking effects are therefore usually limited by the intrinsic properties of known, stable materials. In recent years, advances in focused ion beam (FIB) fabrication have enabled the nanostructuring of bulk crystals into ultraprecise transport devices[1–4], facilitating the investigation of geometrical effects on mesoscopic length scales. In this work, we expand such explorations into three-dimensional (3D), curvilinear shapes, by sculpting helical nanostructure devices from single crystals of the high-mobility, centrosymmetric magnetic Weyl semimetal $Co_3Sn_2S_2$[5,6]. The combination of the imposed chiral geometry and intrinsic ferromagnetism yields nonreciprocal electron transport[7–9]. The high coercivity results in an anomalous, reversable diode effect remnant under zero applied magnetic field, which is orders of magnitude larger than can be explained by a classical self-field mechanism. We argue the enhancement originates from the high carrier mobility and the resulting quasi-ballistic transport: the conduction electron mean free path approaches the length scale of the curvature, resulting in increased asymmetrical scattering at the boundaries. We further demonstrate the inverse effect of the nonreciprocal transport: the field-free, current-induced switching of the magnetisation. The results establish the vast potential of 3D nanosculpting to explore and enrich the functionality of quantum materials.**

In crystalline solids, noncentrosymmetry plays a role in myriad phenomena, being a requirement for ferroelectric polarization[10] and the Berry curvature of non-magnetic Weyl semimetals[11]. In spintronics, which seeks to develop new technologies based on the interplay and coupling of charge and spin, broken inversion symmetry is a key ingredient in the realization of noncollinear magnetic textures[12,13], the ordinary spin Hall effect and spin-orbit torques[14], and thus many field-free magnetisation switching schemes[15]. In the vast majority of cases, the origin of the inversion symmetry breaking is the chiral or polar crystal structure of the underlying atomic lattice, or the interface in a heterostructure film.

In recent years, three dimensional (3D) magnetic and superconducting nanostructures have been developed, with the aim of realizing high-density, low power devices[16–18]. These are typically fabricated via focused electron beam-induced deposition of amorphous cobalt/tungsten[19,20], or by a combination of two photon lithography to create 3D polymer scaffolds and subsequent film deposition[21]. The resulting structures may be designed to impose inversion symmetry breaking via the geometry[22–24]. However, the variety and crystallinity of the resulting materials is limited, preventing the exploration of geometry in single crystalline quantum materials. Meanwhile, over the last decade, the possibility to microstructure single crystals into precise, mesoscale devices using a focused ion beam (FIB) has been developed[2,25], but these efforts have mostly been concentrated on planar devices, lacking curvilinear features.

In this work, we utilize advanced FIB sculpting methods to fabricate 3D helical devices from a bulk single crystal of the high mobility, ferromagnetic Weyl semimetal $Co_3Sn_2S_2$ (illustrated in Fig. 1a). With this method, the intrinsic properties of the high-quality quantum material are preserved, while the imposed chiral geometry breaks inversion symmetry on the mesoscale. This combination results in new electrical functions which are not exhibited by the underlying centrosymmetric material. Specifically, we realize nonreciprocity: the propensity for noncentrosymmetric mediums to exhibit different transmittance in the forwards or backwards directions[7–9,26]. Any noncentrosymmetric conductor may exhibit nonreciprocal transport when time reversal symmetry is additionally broken, which is typically achieved by the application of a magnetic field[7,27,28]. In the present case, we exploit the time-reversal symmetry breaking of the topological ferromagnetism in $Co_3Sn_2S_2$ in combination with the imposed chiral geometry to realise a switchable, zero-field nonreciprocity.

The magnetic Weyl semimetal $Co_3Sn_2S_2$, which naturally has a centrosymmetric crystal structure (Fig. 1b), is well-known for its large, intrinsic anomalous Hall conductivity[5,6]. Using FIB milling, we fabricated 3D helical-shaped devices from a bulk single crystal, achieving inversion symmetry-breaking geometries with both left-handed (LH) and right-handed (RH) chiralities, as shown in Fig. 1c, (see Methods and Extended Data Fig. 1-3 for an overview of all devices, the sculpting procedure, and a discussion of ion damage). $Co_3Sn_2S_2$ exhibits one of the highest carrier mobilities among semimetallic magnets[29,30], with an estimated mean free path on the order of 0.1 to 1 µm at low temperatures (see Methods). By designing the pitch length of the helix $L$ to be 1 µm, we argue that the conduction electrons may propagate coherently over some portion of the device, and the chiral geometry is therefore relevant to the conductivity. In addition, $Co_3Sn_2S_2$ has a Curie temperature $T_C$ of ~160 K and a uniaxial magnetocrystalline anisotropy along the [0001] crystal axis, which we oriented along the helical axis. For comparison, we fabricated an achiral (ACh) rod-shaped device using a similar fabrication process.

In chiral systems, the asymmetrical current-voltage characteristic typically takes the form $V = R(I + B\gamma I^2)$, where $R$ is the resistance, and the $\gamma$ term introduces the nonreciprocity as an odd function of the magnetic field applied along the helical axis[7,9]. The sign of $\gamma$ is opposite for LH and RH chiralities, as illustrated in Fig. 1d. Under an AC current excitation, the nonlinearity results in DC rectification and frequency doubling effects, and may therefore be probed via the second harmonic of the voltage response (see Methods). Figure 2a shows the magnetoresistance, measured as the first harmonic linear response $R_{xx}^{1\omega}$, of the LH device at 10 K, with an AC current excitation of 40 μA, revealing a positive magnetoresistance and features associated with the field-induced reversal of the magnetisation direction, which may have a Berry phase origin[31].

The nonreciprocal character of the LH, RH and ACh devices are revealed in the antisymmetrised second harmonic response, $R_{xx}^{2\omega}$, shown in Figs. 2b-d (see Methods). The LH and RH devices both exhibit a hysteretic $R_{xx}^{2\omega}$ signal, but with inverted sign due to their opposite handedness. The high field, linear response is due to the time-reversal symmetry breaking of the applied magnetic field, while the spontaneous component corresponds to switching of the [0001] axis magnetisation of $Co_3Sn_2S_2$. This anomalous, field-independent component can be modelled as an additional term with coefficient $\Gamma$ in the voltage response,

$$V = R[I + (B\gamma + \Gamma)I^2].$$

It is natural to speculate that this anomalous term is linked to the anomalous Hall effect, and thus the momentum space Berry curvature, intrinsic to $Co_3Sn_2S_2$. Crucially, the ACh sample shows no $R_{xx}^{2\omega}$ response, confirming the imposed chiral geometry as the origin of the nonreciprocity, and conclusively ruling out other extrinsic contributions such as thermal gradients or contact resistance (raw data shown in Extended Data Fig. 4).

Beyond symmetry arguments, one possible underlying mechanism of the nonreciprocity is the self-field effect[7]: a current $I$ passing through the coil-shaped devices induces a magnetic field, $B_c = \mu_0 IN/L$, where $N/L$ is the number of winds per unit length. This coil field will either add or subtract from the applied magnetic field, depending on the direction of the applied current and device chirality. For a device with any kind of magnetoresistance, this may result in a different resistance for the forward and background current directions – a nonreciprocal effect. We illustrate this for the $R_{xx}^{1\omega}$ data from the LH, $L = 1.0$ μm device, shown alongside the inset in Fig. 2a. In Figs. 2b-c, we also plot the estimated contribution from the self-field effect, which was calculated based on the measured $R_{xx}^{1\omega}$ magnetoresistance, and the estimated $B_c = 50$ μT at 40 μA (see Methods). From the comparison, it is clear that the measured $R_{xx}^{2\omega}$ is larger, by around two orders of magnitude, than the possible contribution from the self-field effect, which has been shown multiplied by a factor of 30 for clarity. In addition, from the temperature dependence of the measured $R_{xx}^{2\omega}$ in the LH device, shown in Fig. 2e, it is clear that at some temperatures the sign of the measured $R_{xx}^{2\omega}$ is also different from that predicted by the self-field effect. The results show conclusively that there is an additional and much more significant nonreciprocal mechanism in play.

To probe the role of the conduction electron mean free path, we fabricated helical devices of different dimensions, resulting in the three devices shown in Fig. 3a: the original LH device with $L = 1.0$ μm, a LH device with $L = 0.5$ μm, and a RH device with $L = 2.0$ μm. Finally, a Hall bar shaped device was fabricated with the easy axis oriented out-of-plane, as shown in Fig. 3b. At high temperature, the resistance of a metal or semimetal is dominated by electron-phonon scattering (and electron-magnon scattering for magnetic materials close to $T_C$). However, at low temperatures, these contributions are frozen out, and impurity scattering dominates. Usually the mean free path is smaller than the system size, and the transport is in a diffusive regime. However, in systems where the mean free electron path approaches the system dimensions, at low temperatures we may reach a ballistic or quasi-ballistic transport regime, and the conduction electrons may conceivably traverse the medium without scattering from impurities[32,33]. The remaining resistance derives from dissipative scattering at the imperfect boundary, and the size and shape of the system becomes vital[34,35].

The temperature dependence of the resistance $R$ for each device is plotted in Fig. 3c, normalised by the value at 250 K, $R/R(250\text{ K})$. The datasets diverge at low temperature, indicating a difference in the scattering processes within each device. Given that the samples were prepared from the same bulk crystal, it is reasonable to assume that the impurity scattering contribution should be essentially equivalent. Therefore, the change in residual resistance may be attributed to the finite size effect and crossover into a quasi-ballistic regime with lowering temperature, which results in a larger boundary scattering contribution in the smaller, curvilinear devices. This is demonstrated by the inset of Fig. 2e, which shows the residual resistivity ratio $R(250\text{ K})/R(2\text{ K})$ increasing as a function of the estimated cross section of the conducting channel, $A$, of each device.

We performed temperature dependent measurements on the $L = 0.5$, 1.0 and 2.0 μm helical devices, and extracted $\gamma$ and $\Gamma$ values in each case (see Methods), which are plotted in Figs. 4a and b, and show the significant enhancement of the nonreciprocity with decreasing device dimensions. Figure 4c shows the calculations of the possible self-field effect contribution to $\Gamma$, which do not show significant differences as a function of sample dimensions, and predict features absent from the experimental data. Experimentally, there is a sign change of both $\gamma$ and $\Gamma$, but occurring at different temperatures – around $T_C$ for the former, while at ~120 K for the latter. This appears to indicate a crossover in the dominant asymmetrical scattering mechanism as a function of temperature.

The self-field effect would be agnostic to the various scattering processes, and would cause all magnetoresistance to be equally as nonreciprocal. However, other microscopic mechanisms can alter the nonreciprocity of each scattering contribution individually. One possibility is an intrinsic asymmetry in the electronic bands[28,36,37], in this case due to the chiral shape and a geometrical spin orbit coupling, along the lines discussed concerning chirality induced spin selectivity in chiral molecules[38,39], but the chiral length scale of the helical devices are perhaps too large to directly impact the spin splitting of the electronic structure. Alternatively, the large anomalous Hall effect will produce a radial current density distribution in the arc-shaped conducting channel of the helix. This current distribution would be shifted towards the inner or outer edge depending on the direction of **I** and magnetisation **M**. The polarity **P** of the charge distribution itself would reverse with **I**, such that **P** × **I** would not change sign, and this therefore cannot directly explain the nonreciprocal transport, nor the $M$ switching discussed below. On the other hand, due to the noncentrosymmetric geometry of the helix, the shift of the current density distribution would result in a difference in the average path length for forwards and backwards charge flow, yielding nonreciprocal resistance. However, this mechanism would be independent of the mean free path, and we would expect a response which is constant with temperature, similar to the intrinsic anomalous Hall conductivity, rather than the steady increase towards base temperature seen experimentally.

Instead, we believe that the enhanced nonreciprocity in our system is driven by the asymmetry at the boundaries, as illustrated in Fig. 4d. Considering one segment of the helix to be an arc-shaped conductive channel, we envisage a geometrical polarity, $\mathbf{n}_{out}$ and $\mathbf{n}_{in}$, at the outer and inner boundaries, respectively. In combination with **M**, this may give rise to a toroidal moment $\mathbf{T}_{out,in} = \mathbf{n}_{out,in} \times \mathbf{M}$, oriented along the current direction[40]. In combination with the spin-orbit coupling, this will give rise to local asymmetrical toroidal scattering at the boundaries of the system. Due to the geometry of the sample, with the outer boundary being longer than the inner, the contributions of the two edges will not cancel, giving rise to a net nonreciprocity. This contribution will increase for the smaller helical devices, as well as at lower temperatures as the mean free path lengthens, and the scattering at the boundary dominates. At high temperatures close to $T_C$, on the other hand, thermal fluctuations will give the magnetisation an intermittent in-plane component, $\delta\mathbf{S}$. In the vicinity of the boundaries, the polarity **n** will give rise to an effective Dzyaloshinskii-Moriya interaction[17,41], which leads to a vector spin chirality $\langle \delta\mathbf{S}_i \times \delta\mathbf{S}_j \rangle$ perpendicular to $\mathbf{n}_{out,in}$ ($\langle \cdots \rangle$ being the thermal average) and hence the electrical magneto-chiral effect stemming from the non-zero $\mathbf{M} \cdot \langle \delta\mathbf{S}_i \times \delta\mathbf{S}_j \rangle$ term[27,42–44]. Again, the curved geometry of the device means that the two contributions from the outer ($\mathbf{n}_{out}$) and inner ($\mathbf{n}_{in}$) boundaries will not cancel, giving rise to a net nonreciprocity from the spin fluctuation-driven magnetochiral scattering.

The difference in the crossover temperature between $\gamma$ and $\Gamma$ is intriguing, and this indicates that anomalous and field linear contributions to the nonreciprocity have nonequivalent microscopic origins. We note that the microscopic behaviour of the conduction electrons in response to $B$ and $M$ are different, with the former giving rise to cyclotron motion, while the second directly yields a transverse velocity via the momentum space Berry curvature associated with the inherent Weyl physics, and this may explain the observed differences. In particular, the sign change of $\gamma$ and $\Gamma$ in Figs. 4 **a** and **b** are reasonable, if we can consider that the mechanism for the nonreciprocity changes for the magneto-toroidal scattering ($\mathbf{n} \times M$) to the magneto-chiral scattering ($\mathbf{M} \cdot \langle \delta \mathbf{S}_i \times \delta \mathbf{S}_j \rangle$) as temperature increases. This mimics the crossover behaviour observed in the linear response, where the negative magnetoresistance at high temperatures is dominated by thermal fluctuation-driven spin scattering, while at low temperature we observe the positive ordinary magnetoresistance typical of cyclotron motion in high mobility systems (see Extended Data Fig. 5).

The combination of broken inversion symmetry and time-reversal symmetry may also give rise to the inverse effect of the nonreciprocal conduction: affording control of the chirality[43,45,46] or magnetisation of the system via the applied current direction. In the present case, the chirality is fixed, but the direction $M$ may be switched by current applied parallel or antiparallel to the device, analogous to a spin-orbit torque[14]. By applying positive and negative current pulses of 200 μA to the $L = 0.5$ μm device at 150 K (Fig. 4b), we demonstrate the toggling of the second harmonic voltage under zero applied field (Fig. 4e), indicating the magnetisation switching, which could be considered a new form of magnetic memory. Given the vicinity to $T_C$, the switching is likely thermally assisted, but comparison to the field-dependent data indicates that 100% of the magnetisation reverses (Extended Data Fig. 6).

Both the current-induced switching effect and the anomalous contribution to the nonreciprocity $\Gamma$ are notable, and are the result of the combination of the ferromagnetism of $Co_3Sn_2S_2$, the high mobility of its conduction electrons, and the coupling of both to the imposed chiral geometry. While the appearance of $\Gamma$ may be unique so far, the maximum value of $\gamma = 0.11$ $A^{-1}T^{-1}$ can be compared with past examples: being smaller than the nonreciprocity in a polar semimetal, with its intrinsic origin[28], and from the highly nonlinear superconducting diode effect[8,47,48]; while being larger than that seen in natural chiral magnets[2,31,35] and chiral molecular solids[49]. Two comparable geometrical examples include a macroscopic Bi helix ($\gamma \sim 10^{-3}$ $A^{-1}T^{-1}$)[7] and a chiral carbon nanotube ($\gamma \sim 10^2$ $A^{-1}T^{-1}$)[50], with the present example falling between these two in both the magnitude of $\gamma$ and length scale. We emphasize that the switchable nonreciprocity in our chiral $Co_3Sn_2S_2$ helical devices is just one example of the functionality that may be engineered into existing quantum materials, and greatly expands the possibilities at the emerging frontier of 3D magnetism and spintronics. We anticipate vast potential for progress at the convergence of spin and charge-ordered systems and topological states of matter with device geometry and the coherent, ballistic or hydrodynamic[34,51] transport regimes.

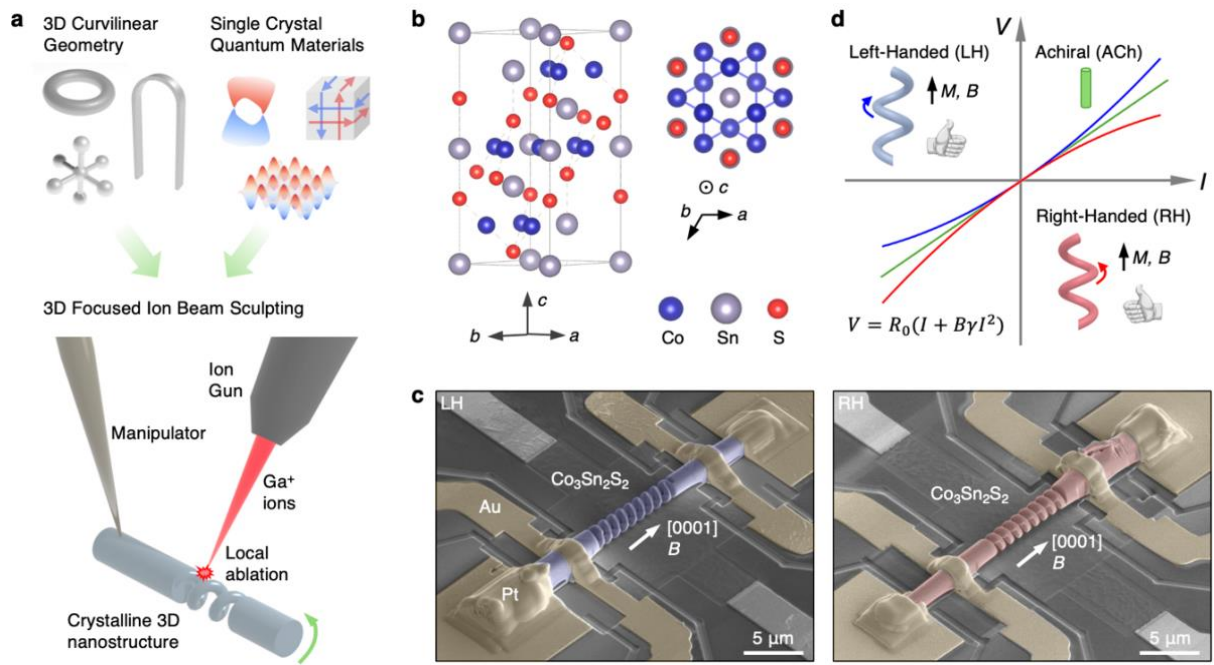

Figure 1 | **Ion-beam sculpting three-dimensional, nanohelix devices of crystalline $Co_3Sn_2S_2$. a** Using focused ion beam milling, the properties of quantum materials may be sculpted into complex 3D curvilinear nanodevices while preserving their intrinsic crystalline properties. **b** The atomic scale, centrosymmetric kagome crystal structure of $Co_3Sn_2S_2$. **d** Schematic illustration of the nonlinear current-voltage (*I-V*) characteristics of left handed (LH) and right handed (RH) conductive helices for a given direction of magnetisation *M* or applied magnetic field *B*. The nonreciprocity is reversed for opposite *M* and opposite chirality. The Achiral (ACh) conductor shows reciprocal *I-V* characteristic. **c** Scanning electron micrographs of nanosculpted $Co_3Sn_2S_2$ helix devices with LH and RH chirality, fixed by Pt welds to Au electrodes patterned on an $Al_2O_3$ substrate. The magnetic field and current were applied parallel to the magnetic easy axis, along [0001]. The scale bars are 5 μm.

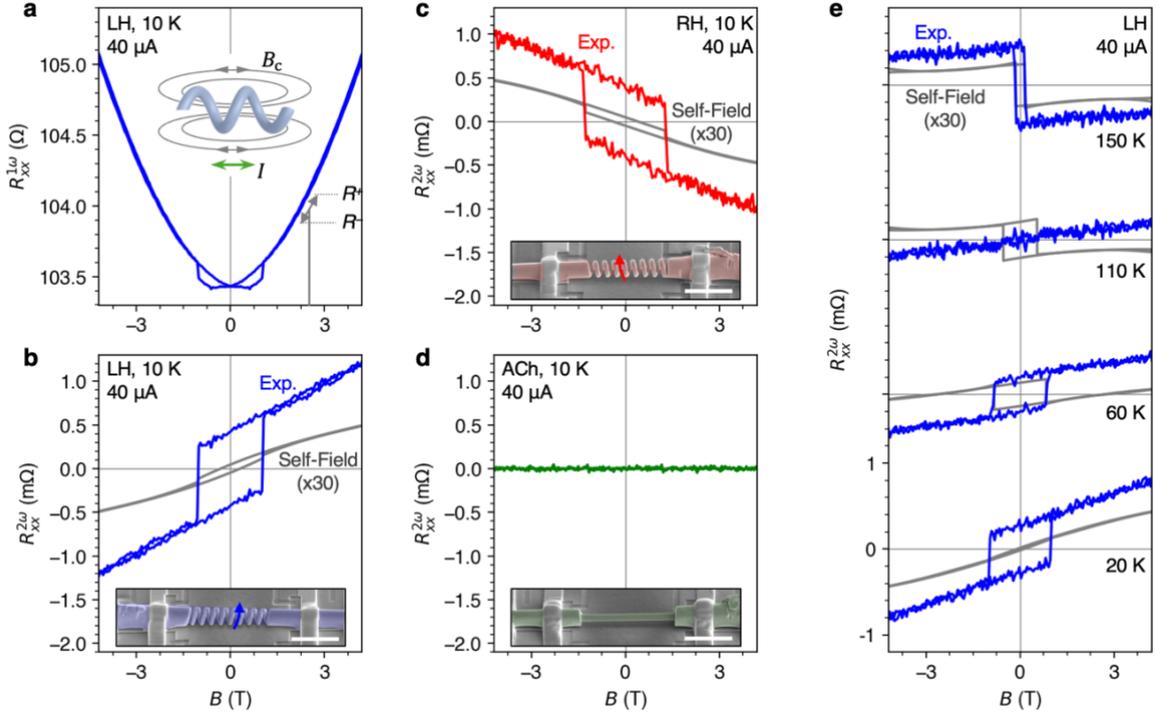

Figure 2 | **Anomalous nonreciprocity in nanosculpted helix devices of $Co_3Sn_2S_2$.** **a** Longitudinal magnetoresistance of the LH helical $Co_3Sn_2S_2$ device with pitch length $L = 1$ μm measured at 10 K with an AC current $I$ of 40 μA. The inset is a schematic illustration of the self-field induced by the coil-like geometry. The self-field adds to or subtracts from the applied $B$ field depending on the current direction, leading to different resistance for the forward and backward directions, labelled $R^+$ and $R^-$. Due to the magnetoresistance, at finite field this may give rise to a contribution to the nonreciprocity with classical origin. **b** The corresponding experimentally measured (Exp., blue) hysteretic second harmonic resistance $R_{xx}^{2\omega}$ at 10 K. The contribution from the self-field effect was estimated from the data in **a**, assuming a coil field of $B_c = \mu_0 I/L$ – approximately 50 μT for this device. The result is plotted in grey, and has been multiplied by a factor of 30 for comparison. **c,d** $R_{xx}^{2\omega}$ measured in the $L = 1.0$ RH device which shows reversed sign; and the ACh device, which shows no measurable nonreciprocity. **e** $R_{xx}^{2\omega}$ measured in the $L = 1$ μm LH device at a range of temperatures. The sign and magnitude of the measured $R_{xx}^{2\omega}$ in comparison to the self-field effect calculation (grey) confirms the nontrivial origin of the nonreciprocity.

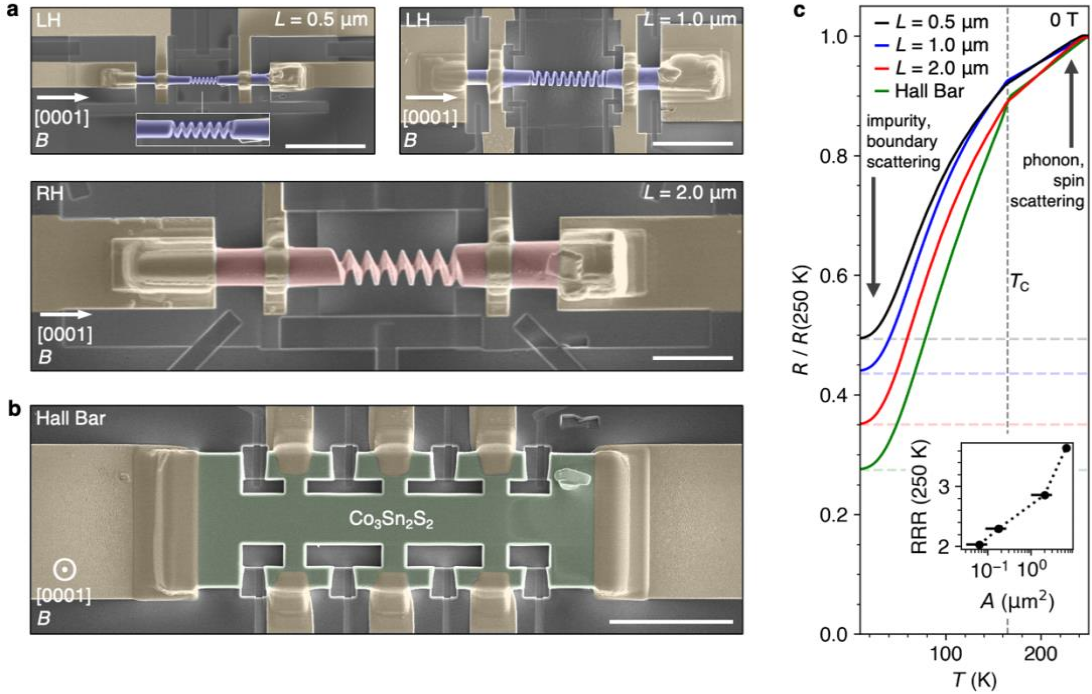

Figure 3 | **Signature of quasi-ballistic transport in $Co_3Sn_2S_2$ devices.** **a** Scanning electron micrographs of further helical $Co_3Sn_2S_2$ devices, with pitch lengths $L = 0.5$, $1.0$ and $2.0$ μm, and chiralities LH, LH and RH, respectively. In all cases, the [0001] axis is aligned along the helical axis. All micrographs of the helical devices are to scale with one another. **b** Scanning electron micrograph of a $Co_3Sn_2S_2$ Hall bar device, with the [0001] axis aligned out-of-plane, perpendicular to the current. All scale bars are 10 μm. **c** The resistance of the $L = 0.5$, $1.0$ and $2.0$ μm helical and Hall bar devices measured as a function of temperature at zero applied magnetic field. The resistance is normalised by the value at 250 K. The vertical dashed line indicates the Curie temperature $T_C$ of the system. Horizontal dashed lines indicate the value approached by each device at base temperature. The inset shows the residual resistivity ratio RRR – $R(250\ K)/R(5\ K)$ – of each device plotted as a function of the estimated cross sectional area of each conductive channel. The resistance at high temperatures is dominated by the phonon and spin/magnon scattering of the conduction electrons, while at low temperatures, impurity scattering and boundary scattering dominate. Since the impurity density should be comparable between the devices cut from the same bulk crystal, a change in RRR indicates a significant contribution from boundary scattering, and thus that the transport is approaching the ballistic regime on the investigated length scales.

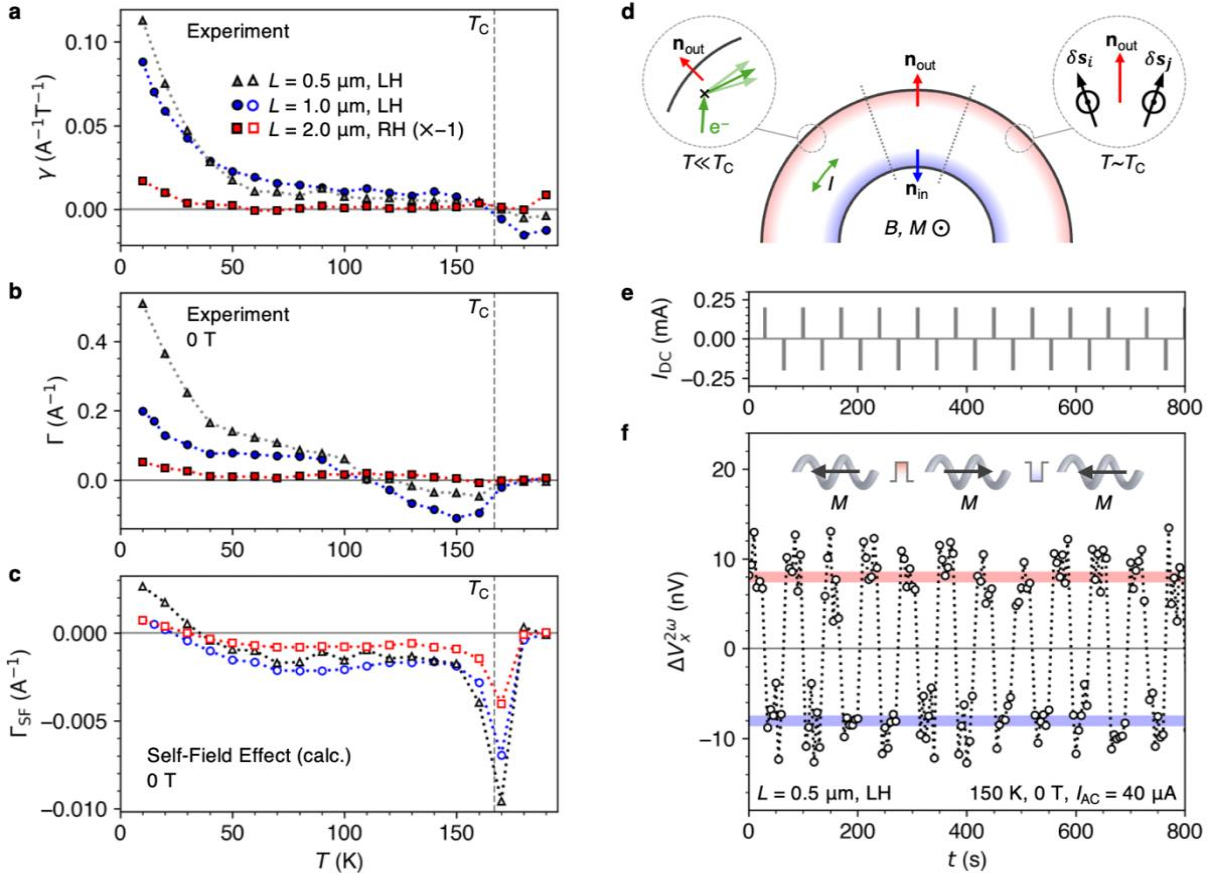

Figure 4 | **Temperature and size dependence of the electrical nonreciprocity and current-induced, field-free magnetisation switching. a,** The fitted value of γ, representing the magnetic field $B$ induced nonreciprocity, for the $L$ = 0.5, 1.0 and 2.0 μm devices, plotted as a function of temperature $T$. **b** The extracted value of Γ, representing the nonreciprocity induced by the spontaneous magnetisation, for the $L$ = 0.5, 1.0 and 2.0 μm devices measured at zero field, plotted as a function of $T$. **c** The estimated value of $\Gamma_{SF}$ from the self-field effect for each device, plotted as a function of temperature. The vertical dashed lines indicate the Curie temperature, $T_C$. **d** Schematic illustration of the asymmetrical scattering mechanisms. We envisage a geometrical polarity **n** at the inner and outer edges of the arc-like conductive channel. At temperatures $T$ far below $T_C$, toroidal scattering at the boundaries will dominate due to the long mean free path. Close to $T_C$, thermal fluctuations may give rise to asymmetrical scattering from transient spin chirality induced by the canting of the spins near to each interface. In both cases, the curved geometry means the contribution from each boundary will not cancel, resulting in a net nonreciprocity. **e** The directional DC current $I_{DC}$ sequence plotted as a function of time $t$. **f** The corresponding response of the $L$ = 0.5 μm LH $Co_3Sn_2S_2$ helical device, showing the measured change in the second harmonic longitudinal voltage $\Delta V_x^{2\omega}$ at 150 K and zero applied magnetic field, plotted as a function of $t$, and measured with an AC current of 40 μA. The repeated toggling of $V_x^{2\omega}$ indicates current-induced switching of the magnetisation within the helical device, illustrated in the inset. The horizontal coloured lines indicate the 100% switching value, as determined from a complementary field sweep measurement (Extended Data Fig. 6).

## Methods

**Sample preparation and device fabrication.** Single crystalline samples of $Co_3Sn_2S_2$ were grown by Bridgman methods. Co, Sn and S were mixed in a stoichiometric ratio and then sealed in a quartz tube. Afterwards, the tube was placed in the Bridgman furnace and heated to 1323 K, and then cooled down to 973 K while moving at a rate of 4 mm per hour. The crystal was oriented using an x-ray Laue camera, followed by cutting and polishing of selected crystal orientations. A dual beam system including Ga ion beam (FIB) and electron beam (Helios 5 UX, Thermo Fisher Scientific) was utilized to fabricate devices from the single crystal via $Ga^+$ ion milling (an overview of all devices is shown in Extended Data Fig. 1, together with their measured $R(T)$ curves). Scanning electron microscopy images of the fabrication process are shown in Extended Data Fig. 2. First, a large slab of $Co_3Sn_2S_2$ material was created by milling trenches in the face of the crystal, which was subsequently picked up via the in-situ micromanipulator (Easy Lift), and fixed to a copper mesh typically used for transmission electron microscopy samples. Subsequent milling processes reduced the thickness of the initial $Co_3Sn_2S_2$ slab to the desired thickness of 1-3 μm. Using ring-shaped milling patterns, the slab was shaped into a comb-like structure, composed of multiple rods with the desired radius. A single rod was then picked up and sculpted by milling with a threading pattern at several rotation angles, creating a 3D helical shape with a hollow center. This final step was performed at a lower ion acceleration voltage of 16 kV to reduce the effect of Ga implantation, which results in a final damaged shell around the crystalline $Co_3Sn_2S_2$ bulk material. The resulting helical-shaped sample was once again picked up with and finally placed onto the $Al_2O_3$ substrate. The Pt deposition system of the Helios 5UX was utilized to fashion contacts between the prepatterned Au electrodes and the $Co_3Sn_2S_2$ helix.

An additional helical shaped sample was fabricated using an identical process, but made to be thin enough to permit transmission electron microscopy measurements. These were able to determine the thickness of the damaged layer, exhibiting a mixed amorphous and polycrystalline structure, to be approximately 4 nm (Extended Data Fig. 3). Assuming a shell of this layer exists around the entire conducting channel, we can estimate the percentage of the cross-sectional area which is damaged to be only between 2 to 5%, increasing for the smaller devices. Thus, for the size of our devices, the transport remains dominated by the more conductive bulk $Co_3Sn_2S_2$ material.

The carrier mean free path $l$ in $Co_3Sn_2S_2$ can be estimated from the semiclassical expression $l = v_F \mu m_{eff}/e$, where $v_F \sim 1 \times 10^5$ ms$^{-1}$ is the Fermi velocity, $m_{eff} \sim 0.6 m_e$ is the rough average effective mass in terms of the electron mass $m_e$, and $e$ is the electron charge. The mobility $\mu$ of $Co_3Sn_2S_2$ has been shown to depend on the growth method of the crystal, with crystalline flakes grown by chemical vapour transport showing mobilities as high as 2500 cm$^2$V$^{-1}$s$^{-1}$, while crystals grown by the Bridgeman method exhibit values around 550 cm$^2$V$^{-1}$s$^{-1}$, with a comparison shown in reference[29]. Thus, we come to a value for $l$ between 0.2 and 0.9 μm, giving an order of magnitude approximation of between 0.1 and 1 μm, which is of the same length scale as the chiral geometry in the devices at low temperatures.

**Transport measurements.** AC transport measurements were performed using a combination of a Quantum Design Physical Property Measurement System (PPMS), a custom-made sample probe, and Stanford Research Systems 860 lock-in amplifiers. The $Co_3Sn_2S_2$ devices were mounted to a sample puck using GE varnish, and contacted with a combination of Au wires and silver paste (Dupont). The puck was mounted to a custom-made probe suitable for use within the PPMS in an in-plane magnetic field geometry, fitted with coaxial cables. With the probe mounted in the PPMS, the applied magnetic field and sample temperature was controlled by the built-in helium cryomagnet. The voltage output of a Stanford Research Systems 860 lock-in amplifier in combination with a load resistor was utilised to apply the AC current to the helical devices with a frequency of 509 Hz. The resulting voltage drop across the devices was measured in a four-probe configuration with the Stanford Research Systems 860 lock-in amplifiers set to acquire the first and second harmonic response. The first harmonic and second harmonic responses were symmetrized and antisymmetrised, respectively, in the typical manner: $R^{sym}(B) = [R_{xx}(+B) + R_{xx}(-B)]/2$, $R^{asym}(B) = [R_{xx}(+B) - R_{xx}(-B)]/2$. Examples of raw data for the three devices featured in Fig. 1 are shown in Extended Data Fig. 4. The high quality of the FIB devices is evidenced by the data measured on the Hall bar device, which reveals longitudinal and Hall conductivities comparable to bulk $Co_3Sn_2S_2$ single crystals, shown in Extended Data Fig. 7, confirming the preservation of the bulk character.

The data measured at each temperature and for each device was fitted in order to calculate the estimated self-field effect contribution, and to extract the values of γ and Γ, with an example shown in Extended Data Fig. 8. To calculate the self-field effect contribution, the first harmonic response $R_{xx}^{1\omega}$ was fitted with a polynomial function up to fourth order in the applied magnetic field. The maximum solenoid magnetic field of each device with helical pitch length *L* for a given current *I* was estimated for each device as $B_c = \mu_0 I/L$. With this estimation, the fitted function was utilised to calculate the possible self-field effect contribution using a functional approach, $R_{xx,SF}^{2\omega}(B) = \frac{1}{4}[R_{xx}^{1\omega}(B + B_c) - R_{xx}^{1\omega}(B - B_c)]$.

The second harmonic response of all the helical devices can be modelled with two contributing terms, $R_{xx}^{2\omega} = \frac{1}{2}R_0\gamma BI + \frac{1}{2}R_0\Gamma I$. Therefore, to calculate γ, the antisymmetrised $R_{xx}^{2\omega}$ data at high field was fitted with a linear trend to extract the gradient, while the zero-field value (y-axis intercept) was extracted to calculate Γ. Current dependent measurements confirmed the origin of the nonlinear 2ω voltage response to be an $I^2$ term, shown in Extended Data Fig. 9 and 10. A linear fit of the current dependence of the ratio $R_{xx}^{2\omega}/R_{xx}^{1\omega}$ yields a more precise Γ value for each device, with a maximum value of 0.7 $A^{-1}$ found for the *L* = 0.5 μm helix, plotted as an inset in Extended Data Fig. 10. For the switching measurements, the DC output voltage capability of the Stanford Research Systems 860 lock-in amplifier was utilised to apply the DC currents.

**Data availability** All experimental data will be uploaded to a public repository prior to publication.

**Acknowledgements** The authors are grateful for the assistance and guidance of the CEMS Semiconductor Science Research Support Team for the use of cleanroom facilities. We are grateful to members of CEMS as well as H. Isobe, L. Turnbull, C. Donnelly and A. Fernández-Pacheco, and M. Hirschberger for the fruitful discussions. This work was supported in part by JSPS KAKEHNHI (Grant No. 23H05431), the Japan Science and Technology Agency (JST) CREST program (Grant No. JPMJCR20T1), and by the RIKEN TRIP initiative Program (Many-body Electron Systems).

**Author Contributions** MTB, NN, MK and YT conceived the project. YF grew the bulk $Co_3Sn_2S_2$ single crystal. MTB fabricated the focused ion beam devices with support from YLC and XY. MTB performed the transport measurements and data analysis with support from YF, IB, MM, and MK. YLC and XY performed the TEM measurements and analysis. MTB, NN and YT interpreted the data and wrote the manuscript, along with input and contributions from all coauthors.

**Competing Interests** The authors declare no competing interests.

**Correspondence** Correspondence and requests for materials should be addressed to M. T. Birch (email: maximilian.birch@riken.jp), or Y. Tokura (email: tokura@riken.jp).

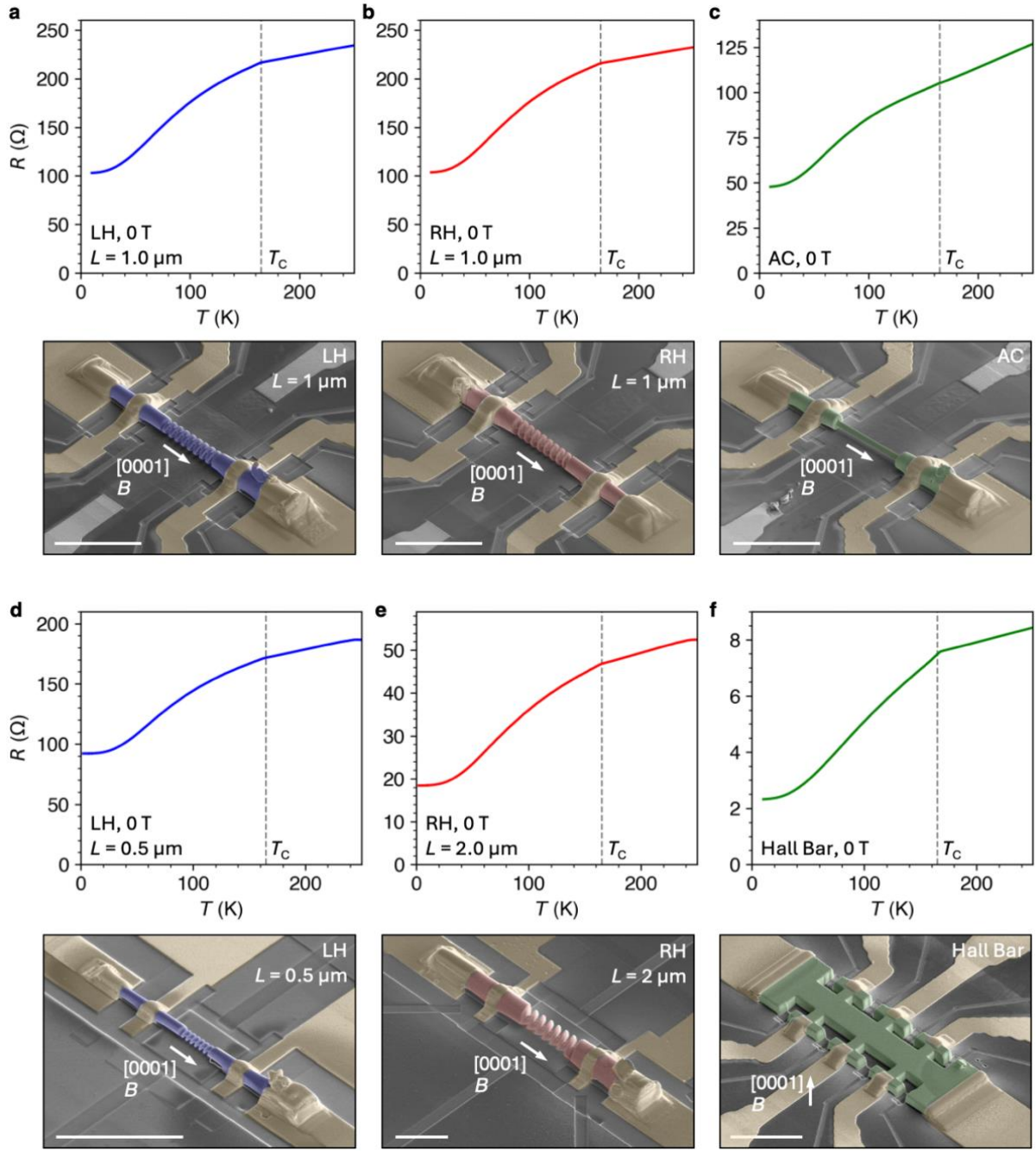

Extended Data Figure 1 | **Focused ion beam sculpted devices of $Co_3Sn_2S_2$. a-f** Resistance measured as a function of temperature, and scanning electron micrographs of the six investigated focused ion beam-fabricated devices: left-handed (LH) helix with pitch length $L$ = 1.0 μm; right-handed (RH) helix with $L$ = 1.0 μm; achiral (AC) rod; LH helix with $L$ = 0.5 μm; RH helix with $L$ = 2.0 μm; Hall bar with thickness 1.6 μm, and width 4.1 μm, respectively. In all cases, the $Co_3Sn_2S_2$ structures were fabricated by Ga ion milling, and fixed to patterned Au electrodes on $Al_2O_3$ substrates with in-situ Pt deposition. All scale bars are 10 μm.

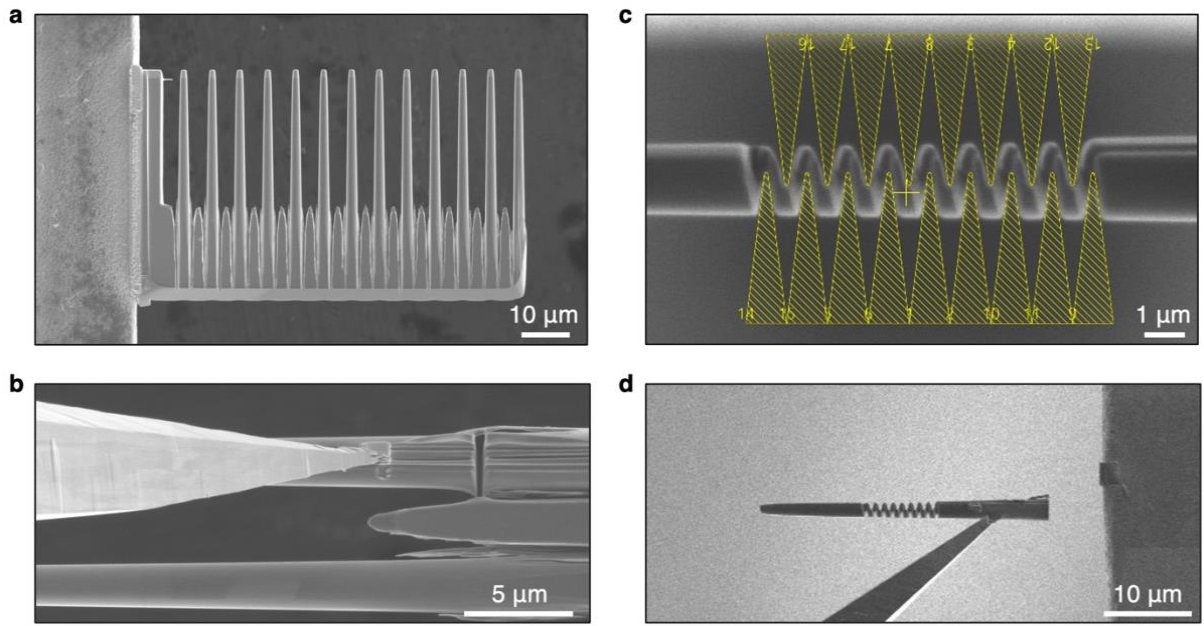

Extended Data Figure 2 | **Co$_3$Sn$_2$S$_2$ helix device fabrication. a** Scanning electron micrograph of comb-like Co$_3$Sn$_2$S$_2$ structure, fabricated from a slab extracted from the bulk single crystal, and attached to the copper grid. The scale bar is 10 μm. **b** A single Co$_3$Sn$_2$S$_2$ rod is picked up by the micromanipulator and transferred to another copper grid. **c** The rod is milled with a thread-like pattern from several angles to create the helical shape. The scale bar is 1 μm. **d** The finished helix-shaped sample is picked up by the micromanipulator and transferred to the Al$_2$O$_3$ substrate and fixed via Pt deposition to the prepatterned Au contacts, creating the final device structure.

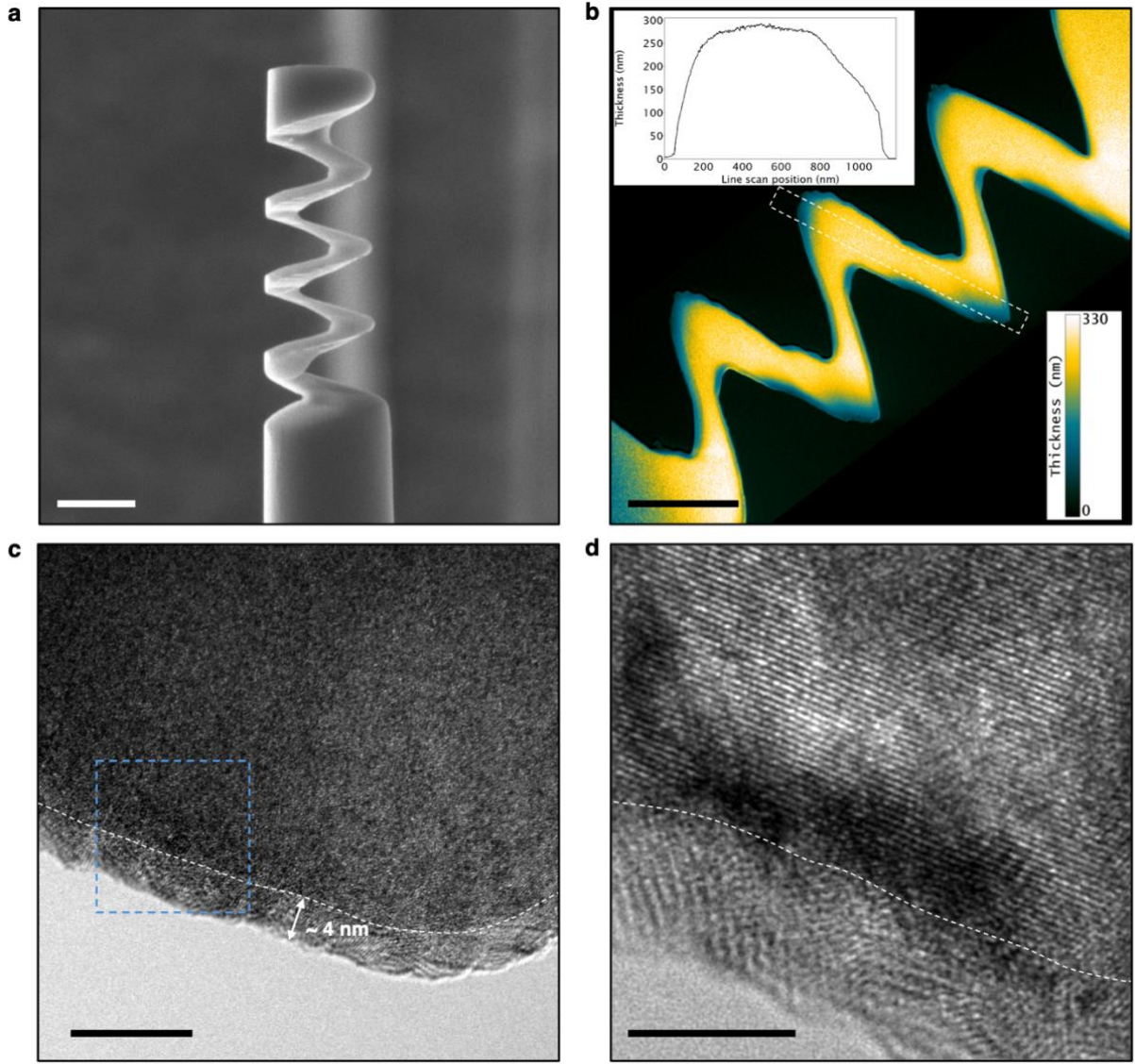

Extended Data Figure 3 | **Transmission electron microscopy analysis. a** Scanning electron micrograph of an additional helical $Co_3Sn_2S_2$ sample, prepared using the identical fabrication process, but made thinner to facilitate the transmission imaging. The scale bar is 1 μm. **b** Thickness map obtained from electron energy-loss spectroscopy (EELS). The inset shows a line profile taken from the top to bottom of the helix, as indicated by the dashed white rectangle, showing a central thickness of ~ 300 nm, tapering to below 150 nm at the edges, making these areas suitable for high-resolution TEM imaging. The scale bar is 500 nm. **c** TEM image at the outer edge of the example spiral. A ~4 nm thick layer consisting of a mixture of amorphous and polycrystalline phases is observed, likely caused by the ion beam damage during FIB fabrication. The scale bar is 10 nm. **d** High-resolution TEM image of the region marked by the blue square in **c**. Clear ordered lattice fringes are observed within the bulk of the helix, confirming the preservation of the single crystalline structure, and indicating that the core material remains largely unaffected by the FIB fabrication. The scale bar is 5 nm.

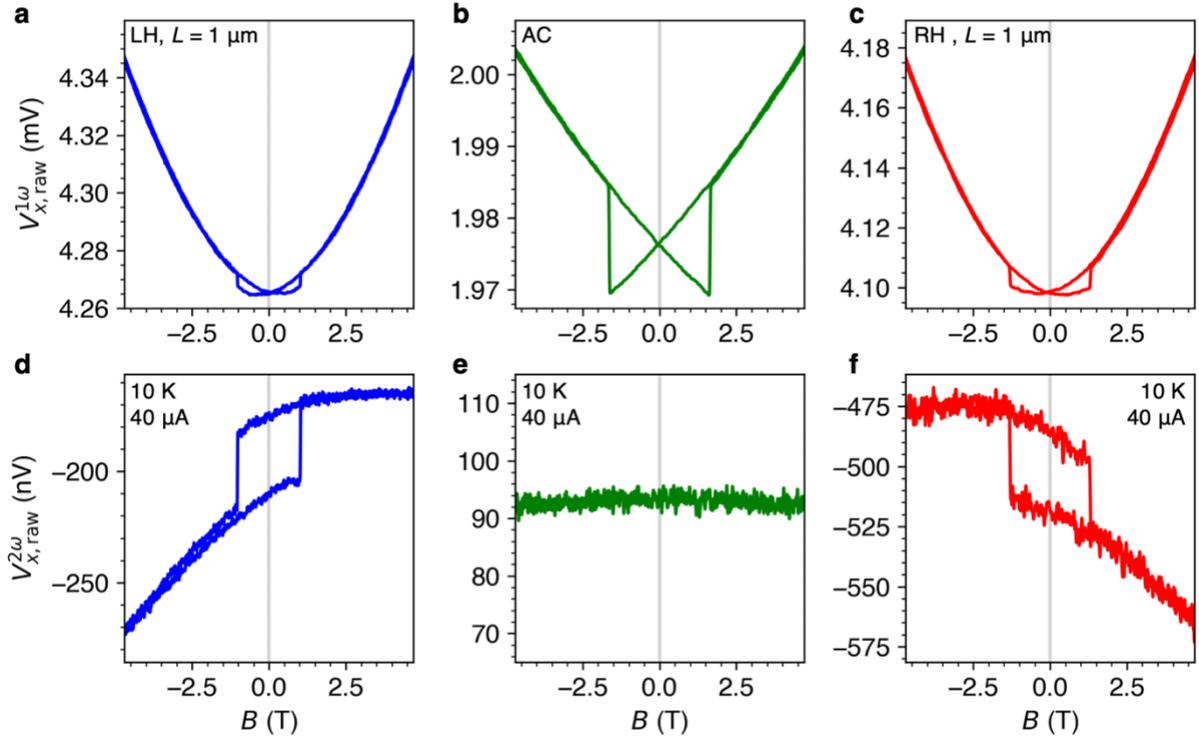

Extended Data Figure 4 | **Example raw voltage data of nonreciprocal $Co_3Sn_2S_2$ devices. a-c** The first harmonic voltage $V^{1\omega}_{x,\text{raw}}$ measured as a function of magnetic field at 10 K, for the left-handed (LH) helix with pitch length $L = 1.0$ μm, achiral (AC) rod, and right-handed (RH) helix with $L = 1.0$ μm. **d-f** The corresponding raw second harmonic voltage $V^{2\omega}_{x,\text{raw}}$ (before antisymmetrisation) measured as a function of magnetic field at 10 K with an AC current of 40 μA.

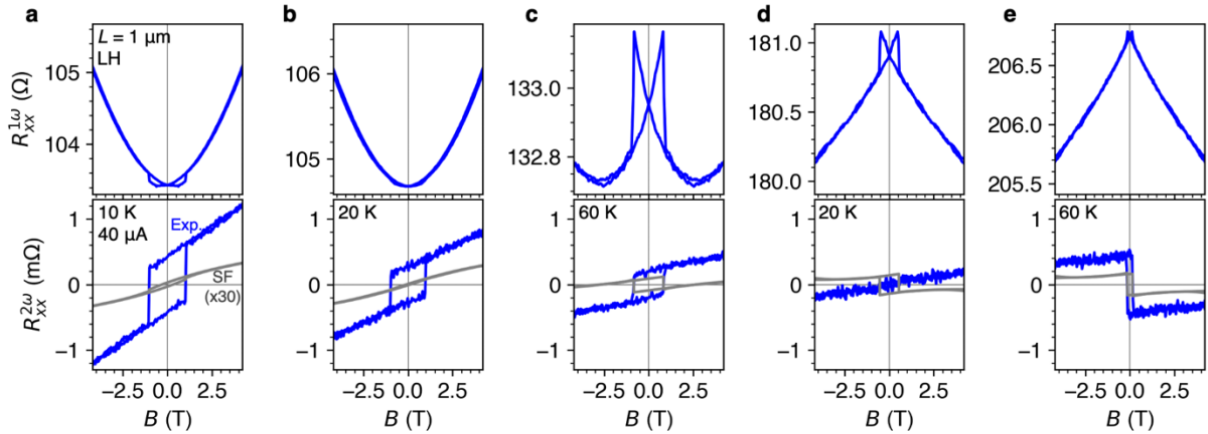

Extended Data Figure 5 | **Crossover in dominant scattering and nonreciprocal scattering mechanisms**. **a-e** The upper panels show the first harmonic resistance $R_{xx}^{1\omega}$ measured as a function of magnetic field, for the left-handed (LH) helix with pitch length $L = 1.0$ μm, at a range of temperatures and an AC current or 40 μA. The lower panels show the second harmonic resistance $R_{xx}^{2\omega}$ measured under the same conditions. The result of the self-field estimation is plotted in grey, and has been multiplied by a factor of 30 for comparison. In both cases, there is a crossover behavior as a function of temperature: the scattering and nonreciprocity at low temperature is dominated by the positive, ordinary magnetoresistance, typical of high mobility systems; while at high temperature the temperature-fluctuation spin/magnon scattering, negative magnetoresistance is the main contribution.

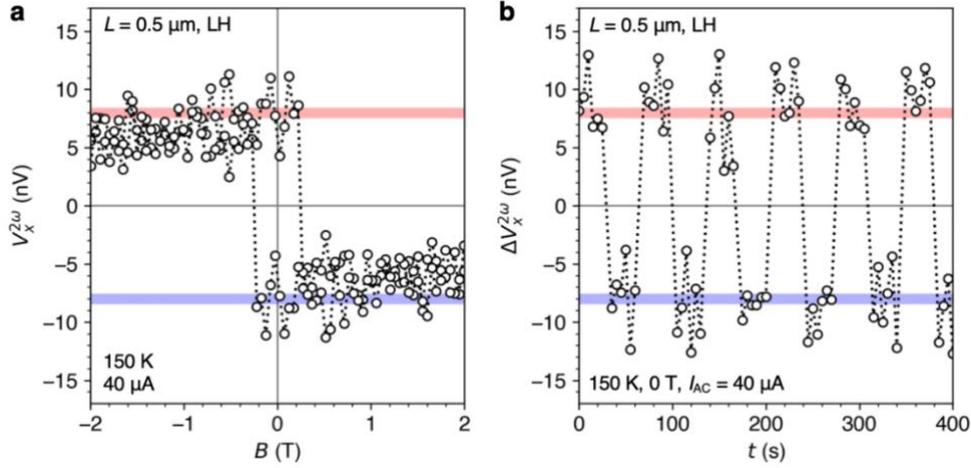

Extended Data Figure 6 | **Current-induced magnetisation switching efficacy**. **a** The second harmonic voltage $V_x^{2\omega}$ measured as a function of magnetic field at 150 K, for the left-handed (LH) helix with pitch length $L$ = 0.5 μm, with an AC current of 40 μA. **b** The response of the device in the current-induced switching experiment, showing the measured change in the second harmonic longitudinal voltage $\Delta V_x^{2\omega}$ at 150 K and zero applied magnetic field, plotted as a function of $t$, and measured with an AC current of 40 μA. In both plots, the horizontal coloured lines are a guide to the eye. Comparison of the change in $V_x^{2\omega}$ indicates that close to 100% of the magnetisation is reversed within the helical device.

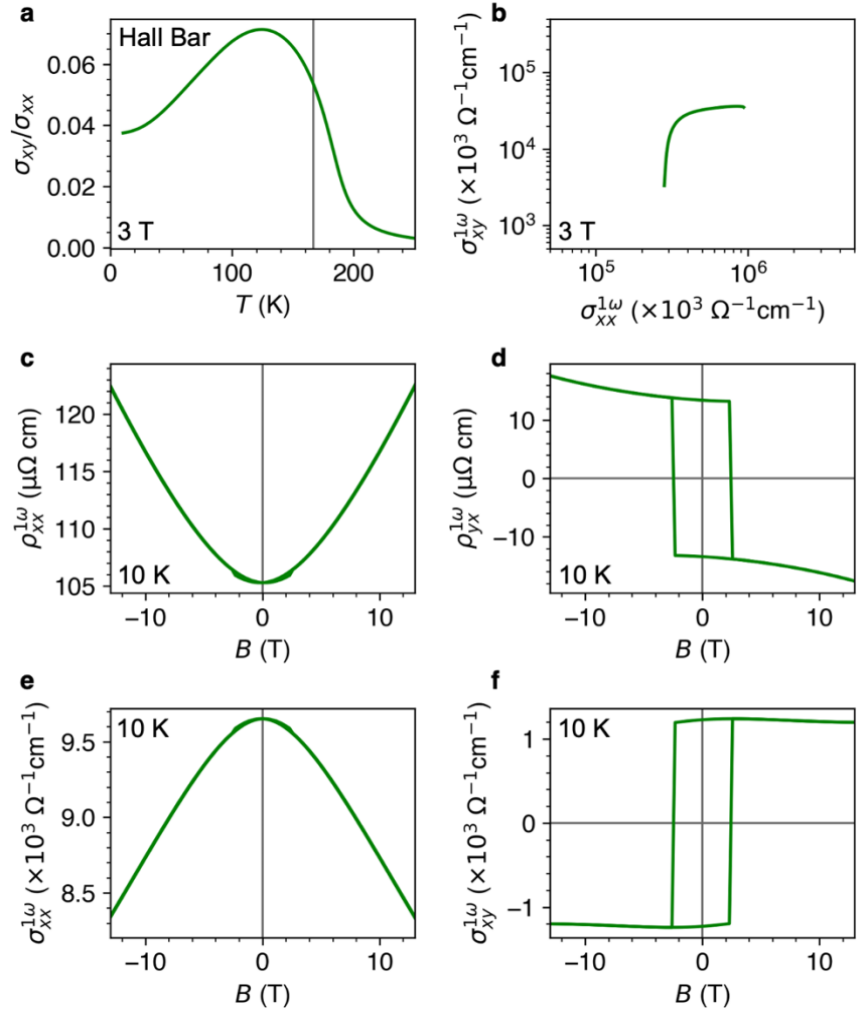

Extended Data Figure 7 | **Characterisation of the $Co_3Sn_2S_2$ Hall bar device. a** Hall angle, calculated as the ratio of Hall and longitudinal conductivities, $\sigma_{xy}/\sigma_{xx}$ measured as a function of temperature under an applied magnetic field of 3 T. **b** From the same temperature dependent data, the Hall conductivity is plotted as a function of the longitudinal conductivity. The constant value of σ$_{xy}$ over a range of temperatures is commonly taken as an indication of the intrinsic nature of the anomalous Hall conductivity, which here is clearly preserved in the device after focused ion beam patterning. **c, d** The longitudinal and Hall resistivities, $\rho_{xx}$ and $\rho_{yx}$ measured as a function of applied magnetic field $B$ at 10 K. **e, f** The corresponding calculated longitudinal and Hall conductivities at 10 K.

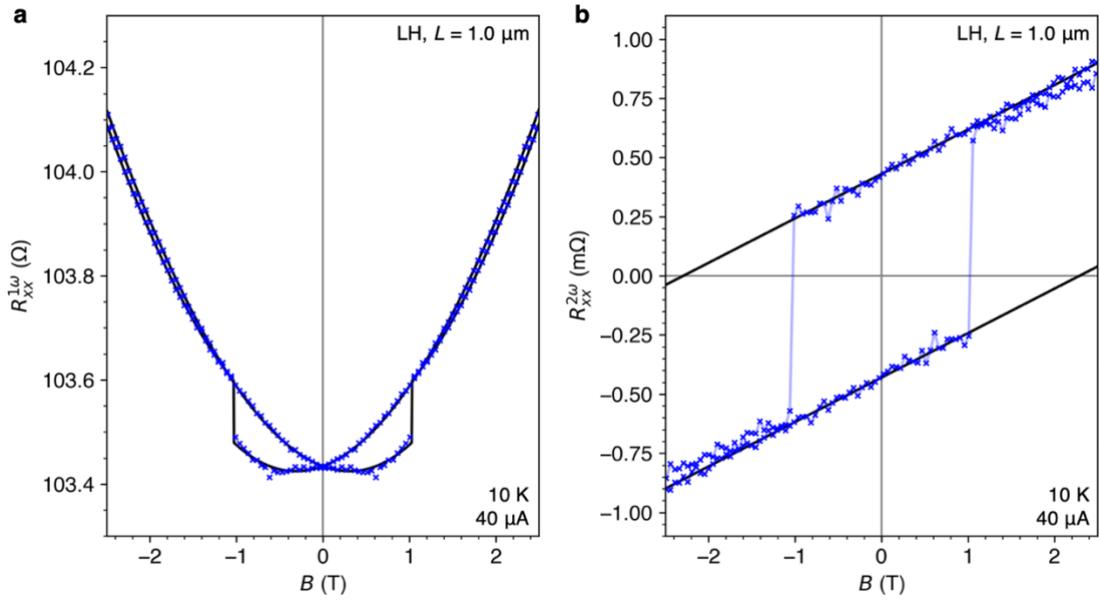

Extended Data Figure 8 | **Fitting of the linear and nonreciprocal responses. a** The longitudinal resistance $R_{xx}^{1\omega}$ measured as a function of applied magnetic field $B$ for the LH, $L$ = 1.0 μm device at 10 K with an AC current of 40 μA. The data was fitted with a polynomial function up to fourth order in the applied magnetic field, yielding the solid black curves. The fit was utilised for the functional calculation of the self-field effect contribution to the nonreciprocity. **b** The second harmonic response $R_{xx}^{2\omega}$ under the same conditions. The high field data was fitted with a linear function to extract the gradient and intercept, for the calculation of $\gamma$ and $\Gamma$. This fitting process was performed for all datasets across all measured devices and temperatures.

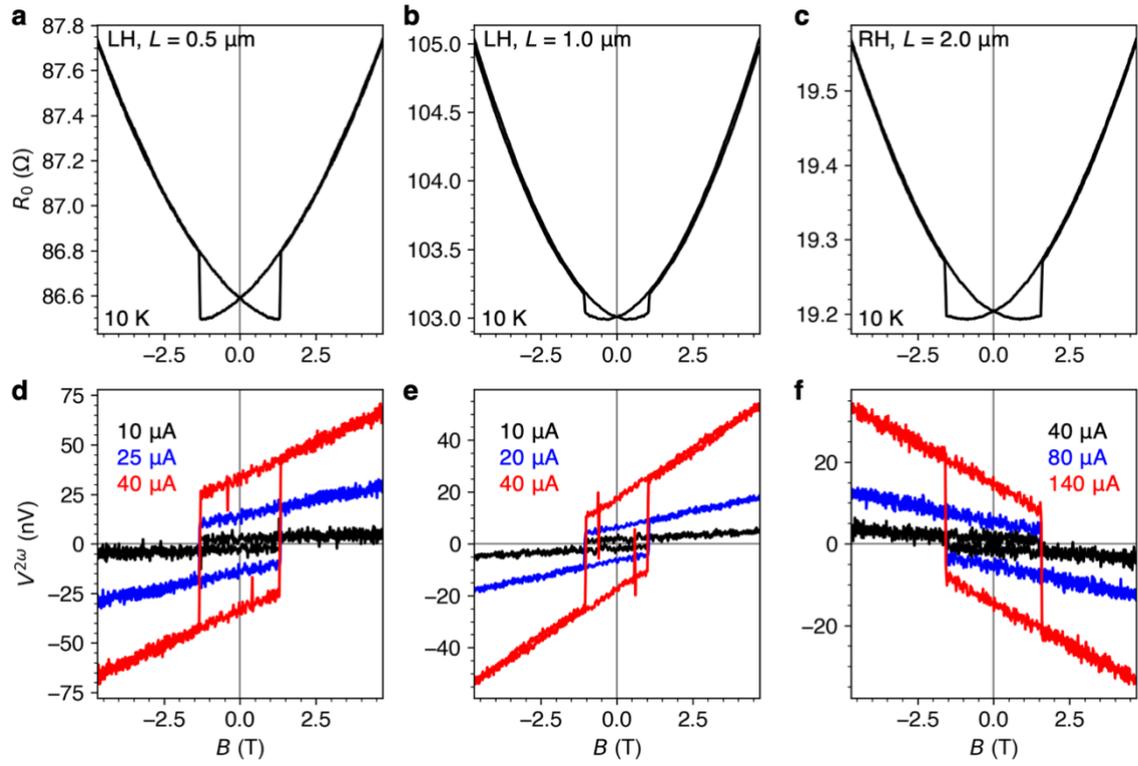

Extended Data Figure 9 | **Current dependent measurements of nonreciprocal helical $Co_3Sn_2S_2$ devices**. **a-c** First harmonic resistance $R_{xx}^{1\omega}$ measured as a function of magnetic field $B$ at 10 K with an AC current of 10 μA, for the helical devices with pitch length $L$ = 0.5, 1.0 and 2.0 μm. **b** The antisymmetrised second harmonic voltage $V_x^{2\omega}$ measured at 10 K with various AC currents, demonstrating the expected $I^2$ scaling for all three devices.

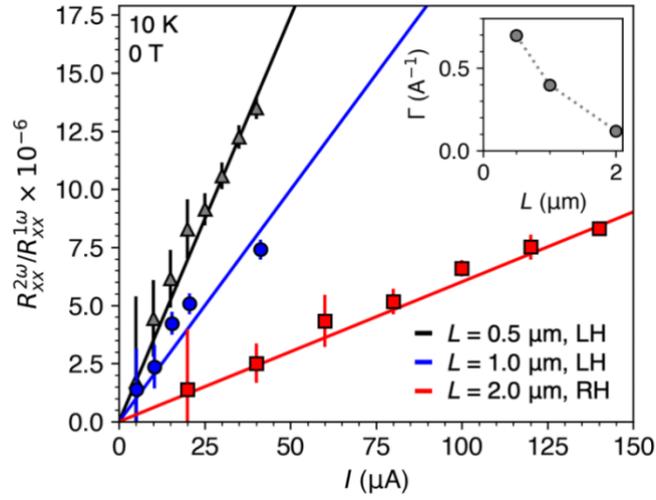

Extended Data Figure 10 | **Fitting the current dependent measurements**. The ratio of the second to first harmonic resistances, $R_{xx}^{2\omega}/R_{xx}^{1\omega}$, at 10 K and zero applied magnetic field, plotted as a function of the applied current $I$ for the three helical devices with pitch lengths $L$ = 0.5, 1.0 and 2.0 µm, as labelled. Error bars indicate the standard error, dominated by the noise in the $R_{xx}^{2\omega}$ measurement. The linear fits give a value for $\Gamma$ of each device via the gradient, which is plotted as a function of $L$ in the inset.